**Favorable modifications of Scrape-Off Layer (SOL) heat flux width through pulsed fuelling in ADITYA-U Tokamak**

SK Injamul Hoque[1,2], Harshita Raj[1,2], Ritu Dey[3], Soumitra Banerjee[1,2], Komal[1,2], Kaushlender Singh[1,2], Suman Dolui[1,2], Ankit Kumar[1,2], Ashok Kumawat[1,2], Bharat Hegde[1,2], Sharvil Patel[4], Kiran Patel[1], Rohit Kumar[1], Suman Aich[1,2], Pramila Gautam[1], Umesh Nagora[1,2], Asha N Adhiya[1], K. A. Jadeja[1], K.M. Patel[1], Ankit Patel[1], R.L. Tanna[1], and Joydeep Ghosh[1,2]

[1]Institute for Plasma Research, Gandhinagar, India
[2]Homi Bhabha National Institute, Mumbai, India
[3]Department of Physics, Indian Institute of Technology Tirupati, Yerpedu, India
[4]University of Virginia, Department of Physics, Charlottesville, USA

Email: injamul.hoque@ipr.res.in

**Abstract:** Enhancement of the scrape-off layer (SOL) heat flux width has been observed in the ADITYA-U Tokamak following the injection of short fuel gas pulses (~$10^{17}$–$10^{18}$ molecules). A notable reduction in parallel heat flux ($q_{||}$) near the last closed flux surface (LCFS) is observed after each pulse. Comparative analysis indicates that pulsed fuelling is more effective in mitigating heat flux with improved core confinement than continuous gas feeding via real-time density control. Analytical and simulation works are also carried out for validation of experimental results. The analytical model shows that SOL width modification cannot be attributed solely to the decrease of temperature due to gas pulse injection; cross-field plasma diffusion also needs to increase. Simulations with the UEDGE code suggest that an increase in both the cross-field diffusion coefficient and inward pinch velocity is necessary to replicate the experimentally observed broadening of the heat flux SOL width. These findings provide insights into efficient SOL heat flux control strategies for future fusion devices.

1.  **Introduction:**

In future tokamak reactors, managing high heat fluxes (~10 MW/m²) will be crucial for maintaining the integrity of plasma-facing components (PFCs) and achieving long-term operation. One key parameter in understanding heat flux profiles in tokamaks is the heat flux SOL width ($\lambda_q$) which represents how quickly the heat flux decreases as it moves from the LCFS to vessel wall, playing a significant role in determining the heat load on the limiter/divertor and other PFCs. Broader $\lambda_q$ is desirable for effective mitigation of the heat load as it increases the weighted area of heat deposition. The International Thermonuclear Experimental Reactor (ITER), like other divertor tokamaks, is designed with start-up and shutdown phases occurring in a limiter configuration. The limiter of ITER is initially designed by assuming a single exponential decay of parallel heat flux ($q_{||}$) [1]. However, infrared (IR) imaging studies at JET reveal the existence of a narrow channel of intense heat flux close to the LCFS, creating a double exponential profile on the inner limiter. This finding suggests that the actual heat loads on ITER's limiter could be up to four times higher than originally predicted [2]. The double exponential behaviour is further discovered in many other tokamaks like KSTAR [3], COMPASS [4], and TORE SUPRA [5]. These works put forward the relevance of studying $\lambda_q$ in the first-wall design of a tokamak. Recent theoretical work reveals that the separation between near and far SOL is due to the presence of strong shear layer [6]. It is well established that $\lambda_q$ decreases monotonically with increasing plasma current and density [7], so it imposes a critical issue for larger tokamaks during the initial ramp-up of plasma current in limiter configuration. Near SOL heat flux is mitigated successfully in TCV with impurity seeding and increased resistivity in SOL due to radiative cooling is found to be the reason [8]. Additionally, divertor heat flux is controlled in real-time with impurity puff using ball-pen and Langmuir probe array as actuator [9]. Recently, it was found in DIII-D that $\lambda_q$ increases with the intensity of high-frequency turbulence in quiescent H-mode plasma [10].

One of the promising solutions to reduce heat load on the divertor is plasma detachment through impurity seeding. But it is often seen that detachment causes severe degradation of core plasma confinement [11]. The integration of detached edge with improved core needs urgent investigation for reactor-relevant tokamak plasma. This has been attempted in TCV, ASDEX and DIII-D tokamaks using advanced divertor configuration [12], using feedback-controlled impurity seeding [13][14]. In ADITYA-U tokamak improvement in core density, temperature is observed after short gas pulse injection due to cold pulse propagation [15]. Several other interesting phenomena are observed in ADITYA-U tokamak due to short gas puff, like mode rotation control of Drift-Tearing Mode (DTM) [16] and sawtooth stabilization [17].

In this paper, results on enhanced heat-flux SOL width and reduced heat load with improved core confinement through short gas pulse injection (pulsed fuelling) are presented. Comparative study on heat flux in the case of pulsed fuelling and continuous fuelling reveals that pulsed fuelling is much more effective in controlling edge heat flux with improved core confinement. Heat-flux SOL width is found to increase periodically with the injection of periodic short gas pulses. An analytical model is introduced to explain the probable reason for the observations. Further, simulation with the UEDGE code is also carried out to estimate the necessary transport dynamics to explain the experimental observations.

The paper is organised as follows: Section 2 presents the experimental setup, Section 3 describes the experimental observations, Section 4 introduces an analytical SOL model for an explanation of experimental observation, Section 5 presents the simulation results, followed by a conclusion in Section 6.

## 2. Experimental setup:

The experiment of heat flux SOL-width modification through pulsed fuelling is conducted in ADITYA-U tokamak, an air core medium-sized tokamak with major radius (R) of 0.75 m and minor radius (a) of 0.25 m. The results presented are from ohmically-heated circular plasma with a toroidal belt limiter and a half poloidal limiter in two toroidal locations. The plasma parameter regime of the analysed shots are as follows: toroidal magnetic field ($B_t$) of 1.2 T, plasma current ($I_p$) of 80-160 kA, central chord-averaged plasma density of $(1 - 4) \times 10^{19} \, m^{-3}$, central chord averaged temperature of 200 - 400 eV, edge density of $(1 - 3) \times 10^{18} \, m^{-3}$, the base pressure of $6 - 8 \times 10^{-9}$ mbar and working pressure of $H_2$ gas is $1 - 4 \times 10^{-4}$ mbar. A proportional Integral and Derivative (PID) based real-time plasma control system is used for real-time horizontal position control [18]. Temporal evolution of the plasma parameters of a typical discharge is shown in the figure (1), loop voltage and plasma current are shown in (b), soft X-ray emission and horizonal position measurement of plasma column are shown in (c), measurement of edge plasma density at r = 24.8 is shown in row (d). Vertical green lines present the gas pulse injected in the flat-top of the plasma discharge.

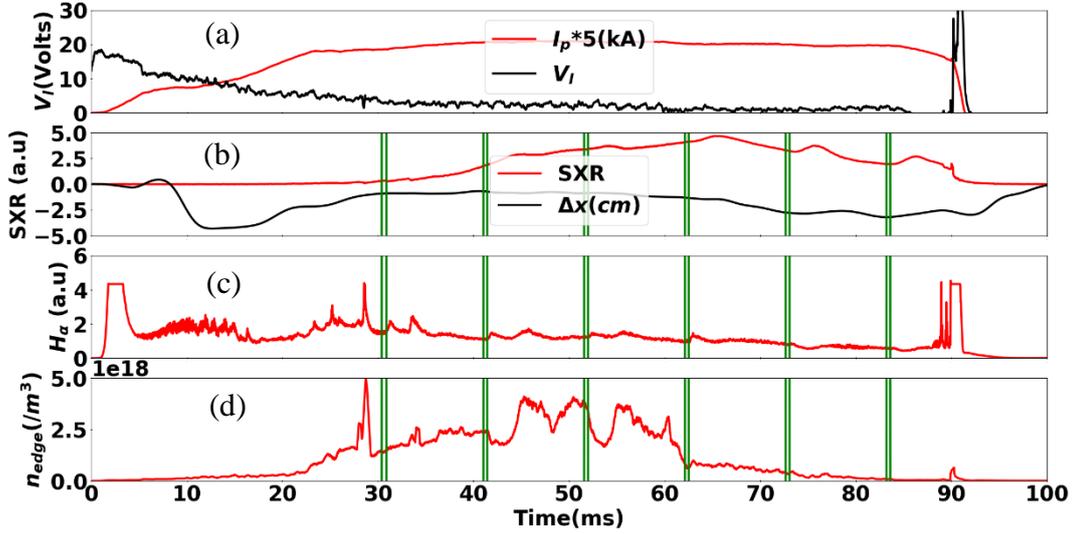

*Figure 1: Typical plasma parameters for shot 36496: (a) loop voltage and plasma current. (b) Soft X-ray and horizontal plasma position, gas pulse (green vertical lines) (c) $H_\alpha$ intensity (d) edge density.*

For the measurement of heat flux and its radial profile, two types of Langmuir probes are used: Triple Langmuir Probe (TLP) and Rake Langmuir Probe (RLP). In case of TLP measurement, one probe is kept floating, and for the remaining two, one is biased with respect to the other. From TLP, heat flux is measured from the simultaneous measurement of plasma density and temperature. RLP is used for the measurement of the radial profile of heat flux. Here, each probe is biased with -120 V (fixed bias) to measure the density profile. It is observed and reported recently that plasma temperature profile in SOL is almost flat in ADITYA-U tokamak [19]. Thus, plasma temperature at each probe location is taken same for heat flux profile is measured from RLP.

A programmable gas-feed system is used for injecting multiple gas pulses of $H_2$ gas in the flat-top of the plasma discharge [20]. For this purpose, prefix pulses of variable pulse-width and voltage are fed to the piezoelectric valve, installed at the bottom port of the vacuum vessel. The amount of $1-10 \times 10^{17}$ molecules are injected into the plasma, which is less than 6% of the background plasma density [21].

For central chord averaged density measurement 100 GHz homodyne microwave interferometer is used. There are also seven heterodyne interferometer chords at r = 0 cm, r = ± 7 cm, r = ± 14 cm, r = ± 21 cm for radial density profile measurement [22]. Two arrays of 16-channel surface barrier AXUV photodiodes are used for Soft X-ray (SXR) monitoring [23]. SXR intensity ratio technique (with two beryllium foil of thickness 10 μm and 25 μm) is employed for estimation of chord-averaged plasma temperature. Spectroscopic measurement consists of photo-multiplier tube (PMT) and interference filters which can give temporal profile of $H_\alpha$, $C^{2+}$, $O^+$ spectral line emissions and visible continuum emission. A toroidal flux loop and Rogowski coil are used for the measurement of loop voltage and plasma current, respectively. All signals are acquired with a 100 kHz sampling frequency except Langmuir probe measurement, for which 1MHz sampling rate is used.

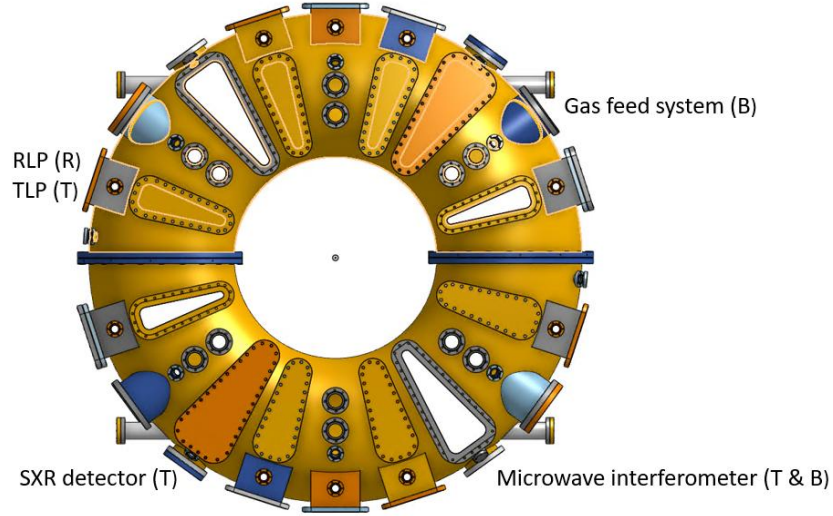

*Figure 2: Schematics of diagnostics position.*

## 3. Experimental observations:

### 3.1 Heat flux measurement with Langmuir probe:

Heat flux is measured by measuring ion saturation current ($I_{sat}$) and temperature ($T_e$) simultaneously with Langmuir probe. Particle flux can be directly determined from the ion saturation current using the relation:-

$$\Gamma_{II} = I_{sat}/(eA) \ldots\ldots (1)$$

Where, A is effective probe area and e is the charge of electron. Parallel heat flux is given by,

$$q_{II} = \gamma \Gamma_{II} T_e \ldots\ldots (2)$$

Here, $\gamma$ denotes the sheath heat transmission factor, which is estimated to be approximately 7 based on an analytical expression that neglects secondary electron emission [24]. Heat flux generally decays exponentially outside the LCFS following the profile –

$$q_{II} = q_{II_0} \exp(-\frac{x}{\lambda_q}). \ldots\ldots (3)$$

Where, $q_{II_0}$ is parallel heat flux at LCFS and x is the distance from LCFS. Exponential fitting the above heat flux profile gives heat flux SOL width, $\lambda_q$. Triple Langmuir Probe (TLP) is used for simultaneous measurement of ion saturation current and temperature with time resolution of 1μs. Each probe has length of 3mm and diameter of 3mm. Figure (3) shows the schematic circuit diagram of TLP. Probe 1 & 2 are biased with respect to each-other with 120V and probe 3 is kept floating. If the applied bias potential ($V_{d2}$) is $\gg T_e$ the temperature can be directly estimated as $T_e = v_{d2}/\ln(2)$. Figure (4) shows the temporal evolution of density, temperature particle flux and heat flux for shot 34655 measured with TLP at $r = 24.4\ cm$.

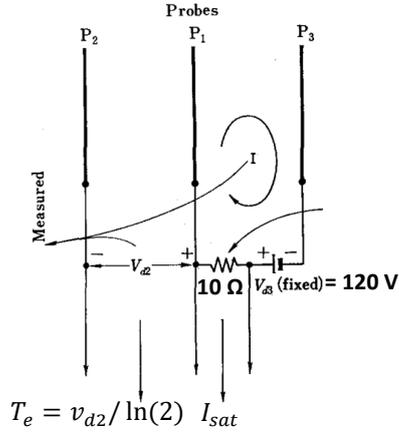
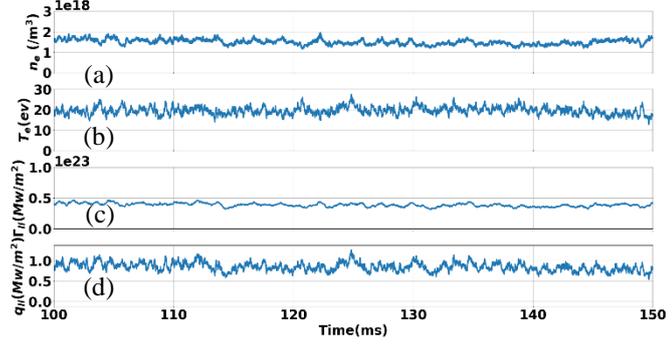

$T_e = v_{d2}/\ln(2)$  $I_{sat}$

Figure 3: Schematics of TLP circuit.

Figure 4: Temporal evolution of edge (a) density, (b) temperature (c) particle flux and (d) heat flux measured by TLP at flat-top of discharge #34655.

## 3.2 Effect of gas fuelling on heat flux:

An experiment is carried out to explore the effect of short gas pulses and continuous gas feeding by real-time density control on edge heat flux. For this, on the flat-top of plasma discharge, pulsed fuelling and continuous fueling are applied in different time windows. For pulsed fuelling, pre-programmed voltage pulses are usually fed to the piezoelectric valve in the flat-top of the plasma discharge. For continuous fuelling, real-time density feedback of measured density from 100 GHz heterodyne microwave interferometry is given to a Field Programmable Gate Array (FPGA) based real-time density control system to maintain constant density [25]. For the real-time density control, a pre-determined density is set, labeled by $n_e^{set}$ as shown in Figure 5. If the measured density ($n_e^{real}$) is less than $n_e^{set}$, the applied voltage in a piezoelectric valve increase, and vice versa. The real-time density control is active in the time window 30-200 ms. After 200 ms, when real-time density control is stopped, a predefined voltage pulses are applied to another piezoelectric valve for maintaining density. A TLP is used for simultaneous measurement of edge density, temperature, and heat flux at r = 24.4 cm. The difference in edge density, temperature, and heat flux for the case of real-time density control and for pulsed fueling is shown in Figure 6. It is observed that in the case of real-time control, there is a steady heat flux of around 1 MW/m² at the edge in the flattop of plasma discharge (100 – 200 ms). Whereas in the case of pulsed fueling, there is periodic suppression of heat flux with a lower level of around 0.2 MW/m², although core density and temperature are well maintained. An increase in core density and temperature after short gas pulse injection has been observed earlier, which is reported as gas puff-induced cold pulse propagation [15]. For further comparison the edge heat flux, core density, and core temperature are averaged in the interval of the gas pulse (7 ms) for 28 ms after and before 200 ms. The time-averaged edge heat flux, core density, and core temperature are plotted in Figure 7 for both the continuous and pulsed fuelling cases. It shows two distinct parameter spaces, one is low edge heat flux and high core density with pulsed fuelling, another is higher edge heat flux with low core density, temperature. This reveals that pulsed fuelling is more effective in controlling edge heat flux with improved core confinement than the real-time density control in ADITYA-U tokamak.

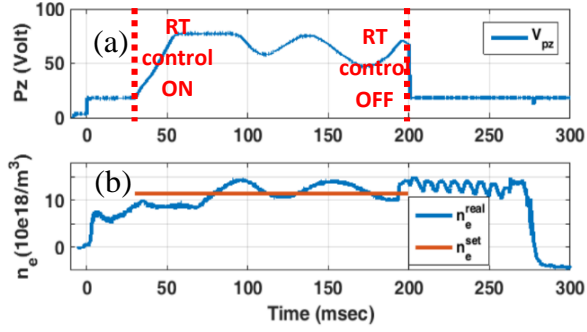

*Figure 5: (a) Voltage applied on piezoelectric valve, the timing of real time control on and off is shown by vertical dotted line. (b) Chord averaged density measured by microwave interferometer ($n_e^{real}$) and predefined density ($n_e^{set}$) for shot 34647.*

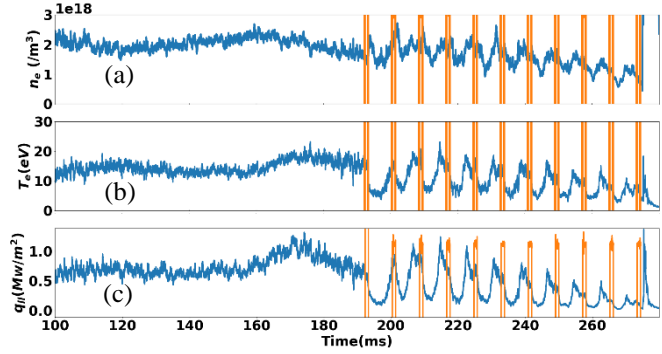

*Figure 6: Measurement of (a) edge density (b) edge temperature and (c) edge heat flux with TLP placed at r = 24.4 cm, vertical lines show the gas pulses.*

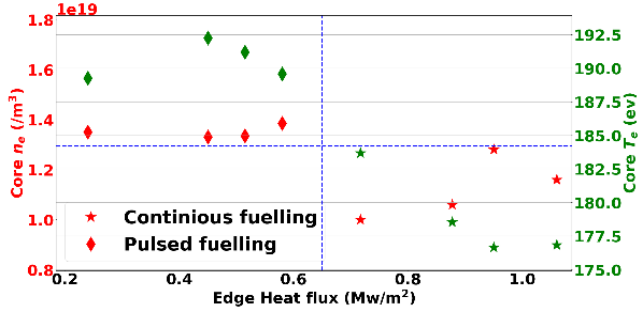

*Figure 7: Time averaged core density (red) and core temperature (green) with edge heat flux (black) for pulsed and continuous fuelling.*

### 3.3 Effect of pulsed fuelling on heat-flux profile and SOL-width:

Periodic short gas pulses are injected in the flat-top of the plasma discharge by feeding voltage pulses of prefixed height and width to the piezo-electric valve. As edge density, temperature decreases, edge heat flux also decreases after the gas pulse injection. Now we are interested in the SOL heat flux profile and SOL width. To study the effect of pulsed fuelling on SOL heat flux profile, a Rake Langmuir probe of inter-probe separation of 8mm, having each probe of length 5mm and diameter of 2.65 mm is used. The probes are biased with a fixed voltage to measure ion saturation current. The temporal profiles of edge density at three locations with gas pulse are shown in Figure 8(a). A periodic decrease in density at the first probe (inside LCFS) and a slight increase in the other probes are observed after each gas pulse. Radial profile of density is plotted in Figure 8(b) for before (40 ms) and after gas puff (43 ms). The data points are averaged over several gas puff pulses, and the error bar is found from the standard deviation of the data points before and after the gas puff. After each gas pulse, the SOL density profile becomes flatter. As in ADITYA-U tokamak temperature does not vary much in the SOL [19], a flat temperature spatial profile with 10 eV in SOL is taken for heat flux profile measurement. Thus, heat flux is obtained from the density profile using equation (2). So, flattening of the density profile causes a flatter heat flux profile after the gas pulse. The heat flux SOL width is calculated from exponential fitting the equation (3). The error bar in the heat flux SOL width measurement is calculated from the residue of the SOL heat flux profile fitted curve. The change of the heat flux profile due to the gas pulse injection is shown in Figure 9(a). The corresponding change in heat flux SOL width is shown in Figure 9(b). It shows SOL width increases by a factor of 1.87 after of gas pulse and decreases thereafter.

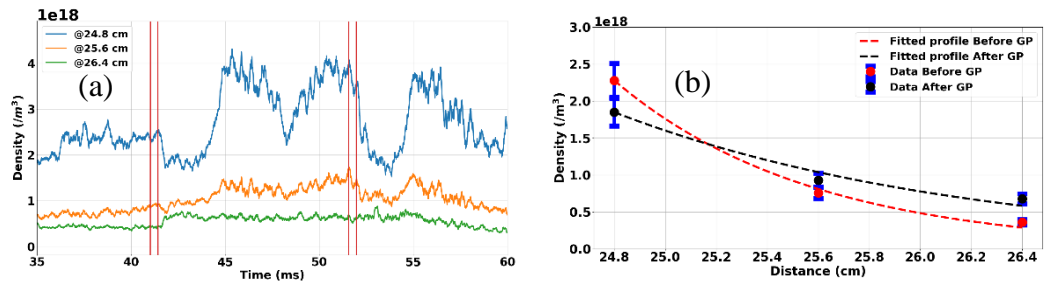

*Figure 8: (a) Temporal profile of edge density at three locations as given in legend with gas pulse (red vertical lines.) (b) Change in radial profile of density after gas pulse is shown, blue and red dots show data points before and after gas pulse respectively, dotted line gives exponential best fit curves and red vertical dotted lines indicates position of LCFS.*

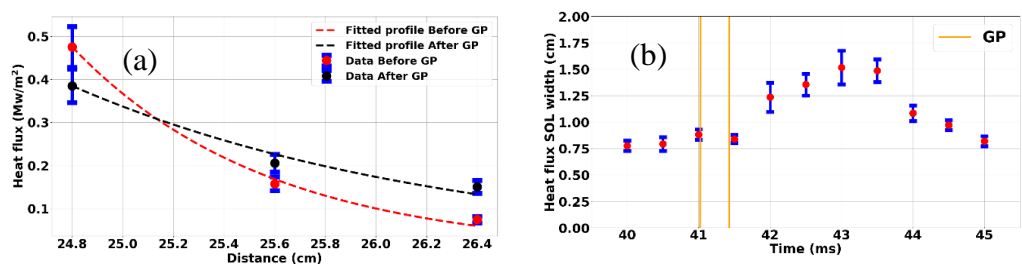

*Figure 9: (a) Heat flux profile before and after gas puff. (b) Evolution of heat flux SOL width with gas pulse.*

## 4. Analytical model for SOL width estimation:

Cross-field diffusion of plasma from the confined region to the SOL is a crucial parameter that controls the SOL density profile. It is expected to have a change in cross-field diffusion after the injection of neutral particles into the plasma. We can model the boundary of the tokamak plasma assuming a finite cylinder. The source is taken only at the Last Closed Flux surface (LCFS) and diffuses into Scrape-off Layer. There are some basic assumptions in this model namely, Electron temperature ($T_e$) is uniform, ions are cold ($T_i = 0$), ionization of $H_2$ molecule is neglected i.e ionization of only H atom is considered. Based on the above assumption, we have the continuity equation for electron [26], [27],

$$\frac{\partial n_e}{\partial t} + \nabla.(n_e \boldsymbol{v}) = <\sigma v>_{ion} n_e n_H - <\sigma v>_{rec} n_e n_{H^+} \quad \ldots \ldots \ldots \ldots \ldots \ldots \ldots (1)$$

For $n_e = n_{H^+} = 1 \times 10^{18} /m^3$, $n_H = 1 \times 10^{17} /m^3$, electron temperature, $T_e = 10 ev$,

$<\sigma v>_{ion} n_e n_H = 5 \times 10^{20} \, m^{-3} s^{-1}$, $<\sigma v>_{rec} n_e n_{H^+} = 5 \times 10^{16} \, m^{-3} s^{-1}$.

So, we can neglect the second term of right-hand side in equation (1). Thus, in steady state we write the equation (1) in the form

$$\nabla.(n_e(\boldsymbol{v}_{ll} + \boldsymbol{v}_\perp)) = <\sigma v>_{ion} n_e n_H \quad \ldots \ldots \ldots \ldots \ldots \ldots \ldots (2)$$

We have the perpendicular diffusion equation, $n_e \boldsymbol{v}_\perp = -D_\perp \nabla n_e \quad \ldots \ldots \ldots \ldots \ldots \ldots \ldots (3)$

From equation (2) and (3) we can get the second order differential equation:

$$\frac{\partial}{\partial r}\left(D_\perp \frac{\partial n_e}{\partial r}\right) + \frac{D_\perp}{r}\frac{\partial n_e}{\partial r} - \frac{n c_s}{q\pi R} + <\sigma v>_{ion} n_e n_H = 0 \quad \ldots \ldots \ldots \ldots \ldots \ldots \ldots (4)$$

If $D_\perp, c_s, n_H$ are independent of position the equation (4) has the simple solution [28]:

$$n_e = n_{e0} K_0\left(\frac{r}{\lambda_n}\right) \quad \ldots \ldots \ldots \ldots (5)$$

Where, $K_0$ is zeroth order modified Bessel function, $\lambda_n$ is density SOL width. For $\frac{r}{\lambda_n} \gg 1$ equation (5) has exponentially decaying solution of the form:

$$n_e = n_{e0} exp\left(-\frac{r_0 - r}{\lambda_n}\right) \quad \ldots \ldots \ldots \ldots (6)$$

Where, $\lambda_n = D_\perp^{\frac{1}{2}} \left(\frac{c_s}{q\pi R} - n_H <\sigma v>_{ion}\right)^{-\frac{1}{2}} \quad \ldots \ldots \ldots \ldots (7)$

Solution of the equation (4) is obtained taking $D_\perp$ and $c_s$ independent of $r$ because $T_e$ is assumed to be uniform. Thus, density SOL width ($\lambda_n$) is same as heat flux SOL width ($\lambda_q$). For realistic solution of $\lambda_n$, $\frac{c_s}{q\pi R}$ should be greater than $n_H <\sigma v>_{ion}$. If $n_H > \frac{c_s}{q\pi R} \times \frac{1}{<\sigma v>_{ion}}$ the right-hand side of Equation (7) becomes positive, leading to increase in plasma density with time within the SOL. As a result, a steady-state equilibrium cannot be sustained. Thus, from equation (7) we can find the maximum neutral density that can be fed to the plasma, $n_{H,max} = \frac{c_s}{q\pi R} \times \frac{1}{<\sigma v>_{ion}} = 8.75 \times 10^{17} /m^3$.

For $\frac{c_s}{q\pi R} \gg n_H <\sigma v>_{ion}$ equation (7) reduces to $\lambda_n = \left(D_\perp \frac{q\pi R}{c_s}\right)^{1/2}$, a well-known formula for simple SOL model [29]. The SOL width can be altered by varying $D_\perp$ or $T_e$ as indicated by the terms in right hand side of of equation (7). There are three distinct cases through which $\lambda_n$ can be modified.

For the first case, let plasma temperature decreases (take 10 eV to 5 eV) and $D_\perp$ is constant (0.2 $m^2/s$) after gas pulse injection. For this case density profile is constructed from equation (6) by calculating $\lambda_n$ from equation (7). The initial condition ($n_0$) is taken from experimentally measured density from Langmuir probe at 24.6 cm. The constructed density profile is plotted with experimental data points in Figure 10.

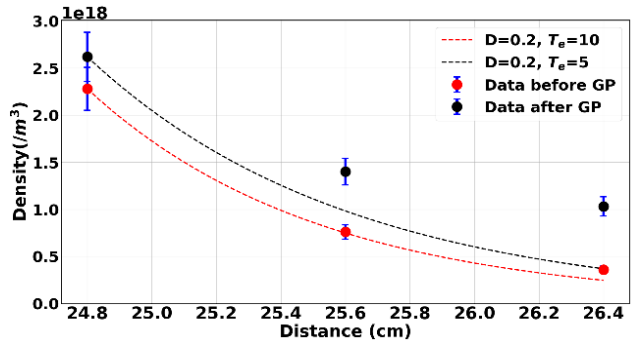

*Figure 10: Constructed density profile (dotted lines) from analytical model with experimental data points (red and black dots) for various $T_e$ and constant $D_\perp$.*

It shows mismatch of theoretical density profile with experimental data in SOL region after gas pulse injection.

For the second case, let perpendicular diffusion coefficient, $D_\perp$ changes from 0.2 to 0.625 $m^2/s$ and plasma temperature, $T_e$ is constant (10 eV). Figure 11 shows the theoretical density profile with experimental data points. In this case modification in $\lambda_q$ is sufficient to match the analytical density profile to the experimentally measured density profile.

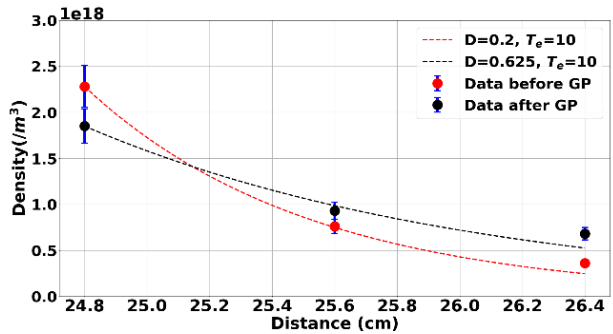

*Figure 11: Constructed density profile (dotted lines) from analytical model with experimental data points (red and black dots) for constant $T_e$, and various $D_\perp$.*

Now let consider the case where both perpendicular diffusion coefficient, $D_\perp$ and plasma temperature, $T_e$ both changes. If we take the change in $D_\perp$ needed for matching experimentally found $\lambda_q$ along with the changes in $T_e$ consisted with equation (7), we found that $D_\perp$ must increases to 0.62 $m^2/S$ from 0.2 $m^2/S$. Figure 12 shows the analytical density profile with $T_e = 10 ev$, $D_\perp = 0.2\ m^2/s$ for before gas puff and $T_e = 5\ ev, D_\perp = 0.62\ m^2/s$ for after gas puff with experimental data points. For this case also analytics profiles matches well with experiment.

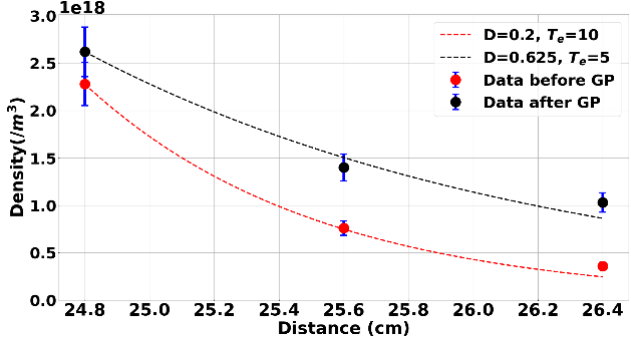

*Figure 12: Constructed density profile (dotted lines) from analytical model with experimental data points (red and black dots) for varying $T_e$ and $D_\perp$.*

So, from the case study, it is concluded that the change in plasma temperature only cannot explain the experimentally observed modification in SOL width. Increase in cross-field diffusion coefficient is required irrespective of temperature change.

## 5. Simulation with UEDGE code:

The UEDGE code [30] uses a fluid transport model to analyze edge-plasma parameters in tokamaks, including plasma density, ion velocity (parallel to the magnetic field), ion and electron temperatures, and electrostatic potential in the edge-SOL region. It incorporates two models for introducing fuel and impurity neutrals: an inertial fluid model and a diffusive model. Furthermore, neutral particles are simulated through a Monte Carlo method, utilizing the DEGAS2 neutral code. The core physical equations are based on Braginskii's formulations, with modifications that account for anomalous or turbulence-enhanced transport across magnetic field lines, while transport along the field remains classical, subject to flux limitations. The $\boldsymbol{E \times B}$ drift is not considered in plasma in this study. The particle source is located at the limiter and is driven only by recycling at the limiter. The detailed incorporation of limiter geometry on the computational mesh and the equations used in the simulation are discussed in the reference paper [31]. In that paper it is shown that edge density profile behavior cannot be explained by a pure diffusive process; a constant inward convection is needed along with cross-field diffusion. For validation of experimental results, simulations of the edge density profile have been done with the UEDGE code, taking the cross-field diffusion coefficient and inward pinch velocity as input. For our study electron and ion density at 2 cm inside LCFS are set to $2.5 \times 10^{18}\ /m^3$, electron and ion temperature ($T_e = T_i$) are set to 40 eV as boundary conditions. Parallel velocity and its derivative are set to zero at this location. For reconstruction of the density profile before gas pulse injection, the cross-filed diffusion coefficient ($D_\perp$) as predicted by the analytical model is taken along with a constant inward pinch velocity. With 0.2 $m^2$/s of $D_\perp$ and inward ware-pinch velocity, Vp of 1.5 m/s simulated profile matches well with the experimental measured density profile. Similar kind of simulation is carried out by M. V. Umansky et al. [32] on edge plasma recycling and transport in the Alcator C-Mod tokamak using the UEDGE code, demonstrating that accurately reproducing the electron density in the SOL at a mid-plane gas pressure of 0.025 mTorr requires a spatially dependent cross-field diffusion coefficient, $D_\perp$, ranging from approximately 1.0 to 0.1 m²/s. Modelling studies of the DIII-D tokamak using the UEDGE code [33] have also revealed the presence of rapid anomalous cross-field plasma transport within the SOL region. Under various plasma conditions in DIII-D, incorporating a peak convective velocity between 50 and 100 m/s, alongside a diffusion coefficient ranging from 0.125 to 0.3 m²/s, resulted in good agreement between simulated edge plasma behavior and experimental measurements in both the SOL and divertor regions. In addition to the previously mentioned studies, UEDGE has also been utilized to model the edge plasma in ITER under a limiter configuration [34]. The results indicate that the extent of limiter penetration significantly influences the distribution of total heat flux between the divertor targets and the outer mid-plane surfaces.

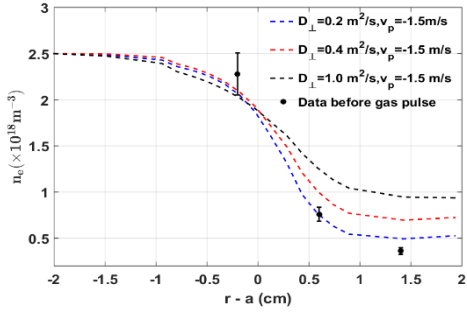

*Figure 13: Simulated density profile with $D_\perp = 0.2$ $m^2/s$ (blue dotted line), $D_\perp = 0.4$ m²/s (red dotted line), $D_\perp = 1$ $m^2/s$ (black dotted line) and inward pinch velocity ($v_p$) of 1.5 m/s with experimental data points before gas pulse.*

Now following the gas puff, when the density profile is simulated using with $D_\perp$ = 0.2 m²/s and Vp = -1.5 m/s (identical to pre-gas puff conditions) the simulated profile fails to align with experimental observations. Increasing the diffusivity to $D_\perp$ = 0.4 m²/s while keeping Vp = -1.5 m/s improves the agreement outside the last closed flux surface (LCFS), but discrepancies persist inside the LCFS. Further increasing $D_\perp$ beyond 0.4 m²/s with the same pinch velocity of Vp = -1.5 m/s only worsens the mismatch. The best agreement with the experimental profile is achieved when $D_\perp$ = 0.6 m²/s is combined with a stronger inward pinch velocity of Vp= 5m/s. This indicates that, to reconcile the post-gas puff profile with simulations using the analytically predicted diffusivity of $D_\perp$ ~ 0.6 m²/s an increased inward pinch velocity is required.

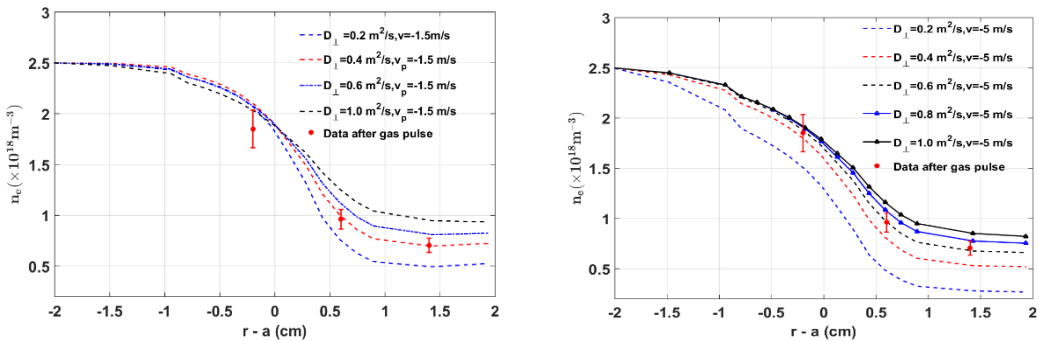

*Figure 14: Simulated density profile with different $D_\perp$ and inward pinch velocity ($v_p$) with experimental data points after gas pulse.*

## 6. Conclusion:

A comparative study of edge heat flux in the case of continuous (through real-time density control) and pulsed (pre-fixed) fuelling is performed in ADITYA-U tokamak. It is found that pulsed fuelling has dual advantage in terms of reduced edge heat flux and improved core confinement. The study on radial profile of edge heat flux during pulsed fuelling shows after each gas pulse edge heat flux profile broadens i.e. heat flux SOL width increases. A 1D analytical model is presented to explain the increased SOL width after gas pulse injection. Analytical results show that decrease in temperature alone due to gas puff cannot explain the increased SOL width. An increase in cross-filed diffusion coefficient also must be incorporated to justify the increase in SOL width. UEDGE code being widely used in many tokamaks to explain density profile behaviour, is employed to study the SOL width modification. Simulation results show that taking cross-field diffusion value same as predicted by analytical model with the inclusion small inward pinch velocity can explain the edge density profile. For the case of density profile after gas pulse injection increase in cross-field diffusion alone with constant inward pinch velocity cannot explain the density profile. Inward pinch velocity must also increase to match the density profile after gas pulse injection. Thus, this enhanced inward pinch [15] is responsible for improved core confinement along with reduced edge heat flux.

In this manuscript, steady state ($\frac{\partial}{\partial t} \sim 0$) behaviour heat flux/density profile (before and after gas pulse injection) is studied through UEDGE code. However, a time dependent simulation will provide more valuable insight on this phenomenon. In future, temporal behaviour of heat flux/density profile during gas pulse injection will be explored with inclusion of source term.

**References:**

[1] M. Kocan *et al.*, "Impact of a narrow limiter SOL heat flux channel on the ITER first wall panel shaping," *Nuclear Fusion*, vol. 55, no. 3, Mar. 2015, doi: 10.1088/0029-5515/55/3/033019.

[2] G. Arnoux *et al.*, "Scrape-off layer properties of ITER-like limiter start-up plasmas in JET," *Nuclear Fusion*, vol. 53, no. 7, Jul. 2013, doi: 10.1088/0029-5515/53/7/073016.

[3] J. G. Bak *et al.*, "Measurement of inner wall limiter SOL widths in KSTAR tokamak," *Nuclear Materials and Energy*, vol. 12, pp. 1270–1276, Aug. 2017, doi: 10.1016/j.nme.2016.12.001.

[4] J. Horacek *et al.*, "Narrow heat flux channels in the COMPASS limiter scrape-off layer," *Journal of Nuclear Materials*, vol. 463, pp. 385–388, Jul. 2015, doi: 10.1016/j.jnucmat.2014.11.132.

[5] S. Carpentier *et al.*, "Study of heat flux deposition on the limiter of the Tore Supra tokamak," *Journal of Nuclear Materials*, vol. 390–391, no. 1, pp. 955–958, Jun. 2009, doi: 10.1016/j.jnucmat.2009.01.246.

[6] F. D. Halpern and P. Ricci, "Velocity shear, turbulent saturation, and steep plasma gradients in the scrape-off layer of inner-wall limited tokamaks," *Nuclear Fusion*, vol. 57, no. 3, Mar. 2017, doi: 10.1088/1741-4326/aa4eb6.

[7] Loureiro J, Silva C, Horacek J, Adamek J, and Stockel J, "Scrape-off layer width of parallel heat flux on tokamak COMPASS," 2014.

[8] F. Nespoli *et al.*, "Impurity seeding for suppression of the near scrape-off layer heat flux feature in tokamak limited plasmas," *Phys Plasmas*, vol. 25, no. 5, May 2018, doi: 10.1063/1.5023201.

[9] I. Khodunov *et al.*, "Real-time feedback system for divertor heat flux control at COMPASS tokamak," *Plasma Phys Control Fusion*, vol. 63, no. 6, Jun. 2021, doi: 10.1088/1361-6587/abf03e.

[10] D. R. Ernst *et al.*, "Broadening of the Divertor Heat Flux Profile in High Confinement Tokamak Fusion Plasmas with Edge Pedestals Limited by Turbulence in DIII-D," *Phys Rev Lett*, vol. 132, no. 23, Jun. 2024, doi: 10.1103/PhysRevLett.132.235102.

[11] A. Kallenbach *et al.*, "Partial detachment of high power discharges in ASDEX Upgrade," *Nuclear Fusion*, vol. 55, no. 5, May 2015, doi: 10.1088/0029-5515/55/5/053026.

[12] H. Raj *et al.*, "Improved heat and particle flux mitigation in high core confinement, baffled, alternative divertor configurations in the TCV tokamak," *Nuclear Fusion*, vol. 62, no. 12, Dec. 2022, doi: 10.1088/1741-4326/ac94e5.

[13] O. Gruber *et al.*, "Observation of Continuous Divertor Detachment in H-Mode Discharges in ASDEX Upgrade," 1995.

[14] H. Q. Wang *et al.*, "Observation of fully detached divertor integrated with improved core confinement for tokamak fusion plasmas," *Phys Plasmas*, vol. 28, no. 5, May 2021, doi: 10.1063/5.0048428.

[15] T. Macwan *et al.*, "Gas-puff induced cold pulse propagation in ADITYA-U tokamak," *Nuclear Fusion*, vol. 61, no. 9, Sep. 2021, doi: 10.1088/1741-4326/ac189b.

[16] H. Raj *et al.*, "Effect of periodic gas-puffs on drift-tearing modes in ADITYA/ADITYA-U tokamak discharges," *Nuclear Fusion*, vol. 60, no. 3, 2020, doi: 10.1088/1741-4326/ab6810.

[17] S. Dolui *et al.*, "Stabilization of sawteeth instability by short gas pulse injection in ADITYA-U tokamak."

[18] R. Kumar *et al.*, "Real-time feedback control system for ADITYA-U horizontal plasma position stabilisation," Apr. 01, 2021, *Elsevier Ltd*. doi: 10.1016/j.fusengdes.2020.112218.

[19] K. Singh *et al.*, "A multi-purpose reciprocating probe drive system for studying the effect of gas-puffs on edge plasma dynamics in the ADITYA-U tokamak," *AIP Adv*, vol. 15, no. 5, May 2025, doi: 10.1063/5.0253274.

[20] K. Patel *et al.*, "LabVIEW-FPGA-Based Real-Time Data Acquisition System for ADITYA-U Heterodyne Interferometry," *IEEE Transactions on Plasma Science*, vol. 49, no. 6, pp. 1891–1897, Jun. 2021, doi: 10.1109/TPS.2021.3082159.

[21]   R. Jha *et al.*, "Plasma Physics and Controlled Fusion Investigation of gas puff induced fluctuation suppression in ADITYA tokamak Investigation of gas puff induced fluctuation suppression in ADITYA tokamak," *Plasma Phys. Control. Fusion*, vol. 51, p. 17, 2009, doi: 10.1088/0741-3335/51/9/095010.

[22]   P. K. Atrey, D. Pujara, S. Mukherjee, and R. L. Tanna, "Design, Development, and Operation of Seven Channels' 100-GHz Interferometer for Plasma Density Measurement," *IEEE Transactions on Plasma Science*, vol. 47, no. 2, pp. 1316–1321, Feb. 2019, doi: 10.1109/TPS.2018.2890307.

[23]   J. Raval, A. K. Chattopadhyay, Y. S. Joisa, S. Purohit, and P. Kumari, "DEVELOPMENT OF MULTIPURPOSE SOFT X-RAY TOMOGRAPHY SYSTEM FOR ADITYA-U."

[24]   J. Marki *et al.*, "Sheath heat transmission factors on TCV," *Journal of Nuclear Materials*, vol. 363–365, no. 1–3, pp. 382–388, Jun. 2007, doi: 10.1016/j.jnucmat.2007.01.197.

[25]   K. Patel *et al.*, "Real-time density feedback control in ADITYA-U Tokamak," *Journal of Instrumentation*, vol. 17, no. 6, Jun. 2022, doi: 10.1088/1748-0221/17/06/P06004.

[26]   N. Bisai, R. Jha, and P. K. Kaw, "Role of neutral gas in scrape-off layer tokamak plasma," *Phys Plasmas*, vol. 22, no. 2, Feb. 2015, doi: 10.1063/1.4913429/110449.

[27]   N. Bisai and P. K. Kaw, "Role of neutral gas in scrape-off layer of tokamak plasma in the presence of finite electron temperature and its gradient," *Phys Plasmas*, vol. 23, no. 9, Sep. 2016, doi: 10.1063/1.4962844/319711.

[28]   M. Mori, N. Suzuki, and Y. Uesugi, "To cite this article: K Uehara et al," 1979.

[29]   P. C. (University of T. C. Stangeby, *The Plasma Boundary of Magnetic Fusion Devices*. Taylor & Francis Ltd, 2000.

[30]   T. D. Rognlien, M. E. Rensink, and G. R. Smith, "Approved for public release; further dissemination unlimited User Manual for the UEDGE Edge-Plasma Transport Code," 2000. [Online]. Available: http://www.llnl.gov/tid/Library.html

[31]   R. Dey *et al.*, "Effect of convective transport in edge/SOL plasmas of ADITYA-U tokamak."

[32]   M. V. Umansky, S. I. Krasheninnikov, B. LaBombard, B. Lipschultz, and J. L. Terry, "Modeling of particle and energy transport in the edge plasma of Alcator C-Mod," *Phys Plasmas*, vol. 6, no. 7, pp. 2791–2796, 1999, doi: 10.1063/1.873236.

[33]   A. Y. Pigarov, S. I. Krasheninnikov, T. D. Rognlien, M. J. Schaffer, and W. P. West, "Tokamak edge plasma simulation including anomalous cross-field convective transport," *Phys Plasmas*, vol. 9, no. 4, p. 1287, Apr. 2002, doi: 10.1063/1.1459059.

[34]   M. E. Rensink and T. D. Rognlien, "Edge plasma modeling of limiter surfaces in a tokamak divertor con®guration."